\documentclass[journal, twocolumn, 10pt]{IEEEtran}

\usepackage{graphicx}
\usepackage{cite}
\usepackage{amssymb,amsmath}
\usepackage{algorithm}
\usepackage{algpseudocode}
\usepackage{epsfig}
\usepackage{epstopdf}
\usepackage{color}
\usepackage{flushend}

\begin{document}

\title{Sharing the Licensed Spectrum of Full-Duplex Systems using Improper Gaussian Signaling}

\pagenumbering{gobble}

\author{{ Mohamed Gaafar, Osama Amin, Walid Abediseid, and Mohamed-Slim Alouini} \\
\small  Computer, Electrical and Mathematical Sciences and Engineering (CEMSE) Division, \\ King Abdullah University of Science and Technology (KAUST), \\ Thuwal, Makkah Province, Saudi Arabia. \\ E-mail: {\{{mohamed.gaafar, osama.amin, walid.abediseid, slim.alouini\}@kaust.edu.sa} \vspace*{-20pt}}
 \thanks{The work of M.-S. Alouini was supported by the Qatar National Research Fund (a member of Qatar Foundation) under NPRP Grant NPRP 5-250-2-087. The statements made herein are solely the responsibility of the authors.}
}

\maketitle

\begin{abstract}
Sharing the spectrum with in-band full-duplex (FD) primary users (PU) is a challenging and interesting problem in the underlay cognitive radio (CR) systems. The self-interference introduced at the primary network may dramatically impede the secondary user (SU) opportunity to access the spectrum. In this work, we attempt to tackle this problem through the use of the so-called \textit{improper Gaussian signaling}. Such a signaling technique has demonstrated its superiority in improving the overall performance in interference limited networks. Particularly, we assume a system with a SU pair working in half-duplex mode that uses improper Gaussian signaling while the FD PU  pair implements the regular proper Gaussian signaling techniques. First, we derive a closed form expression for the SU outage probability and an upper bound for the PU outage probability. Then, we optimize the SU signal parameters to minimize its outage probability while maintaining the required PU quality-of-service based on the average channel state information. Finally, we provide some numerical results that validate the tightness of the PU outage probability bound and demonstrate the advantage of employing the improper Gaussian signaling to the SU in order to access the spectrum of the FD PU.

\end{abstract}

\section{Introduction}

Cognitive radio (CR) is a promising technology that mitigates the spectrum scarcity which resulted from the recent tremendous growth of wireless devices over the past decade. As many licensed primary users (PU) block the available spectrum, underlay CR system exploits the same spectrum resources and allows secondary users (SU) to coexist with PU without degrading the PU quality-of-service (QoS) \cite{zhao2007survey}.  
CR systems can be incorporated with other communication techniques, such as cooperative  and full-duplex (FD) communications, to improve the spectrum utilization of the communication networks and the SU performance \cite{kim2012optimal, zhongTOAPPEARperformance} .

FD is a spectral efficient paradigm that allows the communication nodes to  transmit and receive simultaneously at the same frequency. FD has recently attracted wide attention especially after the progress in self-interference cancellation, which gives a great promise in practical realization \cite{sabharwal2014band}. In underlay CR systems, FD is used to compensate the spectral efficiency loss of cooperative systems that is used to increase the SU coverage \cite{kim2012optimal, zhongTOAPPEARperformance}. In addition to that, FD is used to achieve simultaneous sensing and data transmission for SU, or possibly receive data from the other SU node during the transmission according to the channel conditions \cite{afifi2015incorporating}. The existing research on CR systems avoided sharing the spectrum with in-band FD PU due to the increased interference limitations at the PU side, which can impede the operation of underlay CR systems. Thus,  investigating communication systems that can relieve the interference signature on the PU while improving the SU performance becomes imperative.

Improper Gaussian signaling has proven its superiority over proper Gaussian signaling to improve the achievable  rate in interference-limited networks  \cite{zeng2013transmit, ho2012improper, zeng2013optimized}. In CR systems,  improper Gaussian signaling is employed in \cite{lameiro2015benefits}, where the PU is assumed to work in half-duplex mode with proper Gaussian signaling. On the other hand, the SU is assumed to use improper Gaussian signaling and have perfect instantaneous channel state information (CSI) of all PU and SU communication links. Improper Gaussian signaling showed better performance than proper Gaussian signaling when the PU is not highly loaded.   

In this paper, we assume the challenging spectrum sharing scenario where PU is working in FD mode and inspect the possibility of inserting the SU into operation without deteriorating the PU QoS. For this purpose, we adopt the improper Gaussian signaling for the SU to create a room for spectrum sharing with the FD PU while satisfying its QoS. To the best of our knowledge, this is the first work that investigates sharing the licensed spectrum of in-band FD PU. Our main contributions are summarized as follows:   
 \begin{itemize}
 \item Derive a closed form expression for the outage probability of the SU employing improper Gaussian signaling and subjected to interference from the FD PU.
 \item Derive a tight upper bound for the outage probability of FD PU, which subjected to SU and self interference,   in terms of the SU signal parameters represented in the SU power and the circularity coefficient. 
 \item Design the SU signal parameters to improve its performance based on the average CSI while maintaining  acceptable PU QoS . In this context, the system performance is measured in terms of the outage probability.
 \item Investigate through numerical results the benefits that can be reaped by employing the improper Gaussian signaling for the SU.
 \end{itemize}

\section{System Model}
\begin{figure}[!t]
\centering
\includegraphics[width=3in]{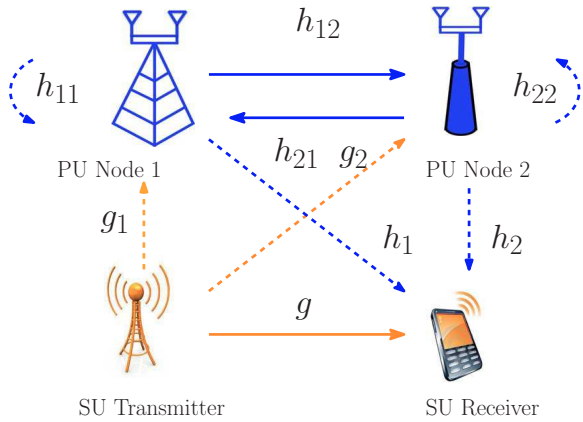}
\caption{System model.}
\label{fig1}
\end{figure}

Consider an underlay CR system, where a half-duplex SU pair coexists with an in-band full-duplex PU pair as depicted in Fig. \ref{fig1}. In this scenario, we have three simultaneous different transmissions occur in the same network over the same frequency. Before proceeding in describing the system components, we define the following terms:

\textit{Definition 1}: The variance and pseudo-variance of a zero mean complex random variable $x$ are defined as $\sigma _x^2 = {\mathbb{E}} {{{\{\left| x \right|}^2\}}} $ and $\hat{\sigma} _x^2 = {\mathbb{E}} {{\{{ x }^2\}}} $, respectively, where $\mathbb{E}\{.\}$ is the expectation operator and $\left| . \right|$ is the absolute value \cite{Neeser1993proper}.

\textit{Definition 2}: The proper signal has a zero  $\hat{\sigma} _x^2 $, while improper signal has a non-zero $\hat{\sigma} _x^2 $.  

\textit{Definition 3}: The circularity coefficient $\mathcal{C}_x$ measures the degree of impropriety of signal $x$ and is defined as $\mathcal{C}_x = \left|\hat{\sigma} _x^2 \right|/\sigma _x^2$, where $0 \le {\mathcal{C}_x} \le 1$. $\mathcal{C}_x=0$ denotes \textit{proper} signal and $\mathcal{C}_x=1$ denotes \textit{maximally improper} signal. 

The PU nodes are assumed to use zero-mean proper Gaussian signals $x_i, i \in {\{1,2\}}$ with a unit variance, while the SU transmitter uses an improper Gaussian signal $x_\mathrm{s}$ with a unit variance and a circularity coefficient $\mathcal{C}_x$. The received signal at the  PU node $j$, where $ j \in {\{1,2\}}$, $i \neq j$, is expressed~as
\begin{equation} \label{pu_sig}
{y_j} = \sqrt {{p_i}} {h_{ij}}{x_i} + \sqrt {{p_j}} {h_{jj}}{x_j} + \sqrt {{p_{\mathrm{s}}}} {g_{j}}{x_{\mathrm{s}}} + {n_j},  
\end{equation}
where $p_i$ is transmitted power of the PU node $i$, ${p_{\mathrm{s}}}$  is the SU transmitted power, $n_j$ is the noise at the  PU node $j$ receiver, $h_{ij}$ denotes the communication channel between PU node $i$  and PU node $j$, $g_{j}$ represents the interference channel between the SU transmitter and PU node $j$ and $h_{jj}$ represents the residual self interference (RSI) channel of node $j$ after undergoing analog and digital cancellation techniques. We assume that the RSI is modeled as a zero mean complex Gaussian random variable as in \cite{kim2012optimal, day2012full}.  As for the PU direct links, we could assume channel reciprocity, i.e., $h_{ij}=h_{ji}$, however it might not be the case when the two PU nodes use different spatial antennas locations or the receivers' front end and transmitters' back end are not perfectly matched \cite{biglieri2007mimo}. Hence, we adopt the general assumption of different forward and reverse PU links. 

In the same time, the SU operates in half-duplex mode and the received signal is expressed as
\begin{equation} \label{su_sig}
{y_{\mathrm{s}}} = \sqrt {{p_{\mathrm{s}}}} g{x_{\mathrm{s}}} + \sqrt {{p_1}} {h_{1}}{x_1} + \sqrt {{p_2}} {h_{2}}{x_2} + {n_{\mathrm{s}}},
\end{equation}
where $n_\mathrm{s}$ is the AWGN at the SU receiver, $h_{i}$ is the interference channel of PU node $i$ on the SU receiver, $g$ denotes the direct channel between the SU transmitter and receiver. The SU transmitter needs to adjust its power in order not to exceed the allowable interference level at the PU receiver. The channels in the described system are modeled as  slow Rayleigh flat fading channels  and the noise at the receivers end is modeled as a zero mean additive white Gaussian noise with variance~$\sigma^2$. The transmitted signals, channel coefficients and noise components are assumed to be independent of each other, except for $h_{ij}$ and $h_{ji}$, which can be dependent on each other.

As a result of using improper Gaussian signaling, the achievable rate for the PU node $i$ is given by  \cite{zeng2013transmit, lameiro2015benefits},
\begin{align}\label{pu_rate}
\!\!\!\!\!\!\!\!\!\!\!\!\!\!\!\!\!\!\!\!\!\!\!\!\!\!\!\!\!\!\!\!\! \!\!\!\!\!\!\!\!\!\!\!\!{R_{{{\mathrm{p}}_i}}}\left( {{p_{\mathrm{s}}},{\mathcal C_x}} \right) = {\log _2}\left( {1 + \frac{{{p_i}{\gamma _{{{\mathrm{p}}_i}}}}}{{{p_j}{\upsilon _{{{\mathrm{p}}_j}}} + {p_{\mathrm{s}}}{{\mathcal I}_{{{\mathrm{s}}_j}}} + 1}}} \right) \nonumber \\  \qquad  \qquad \qquad \qquad \qquad \qquad +\frac{1}{2}{\log _2}\left( {\frac{{1 - {\mathcal C}_{{{{y}}_i}}^2}}{{1 - {\mathcal C}_{{{\mathcal{I}}_i}}^2}}} \right),
\end{align} 
where ${\mathcal C}{_{{{y}_i}}}$ and ${{\mathcal C}_{{{\mathcal{I}}_i}}}$ are the circularity coefficients of the received and interference-plus-noise signals at PU node $i$, respectively, ${\gamma _{{{\mathrm{p}}_{_i}}}} = {{{{\left| {{h_{ij}}} \right|}^2}}} / {{{\sigma ^2}}}$ is the channel-to-noise ratio (CNR) of the PU channel from PU node $i$ to PU node $j$,  ${{\mathcal I}_{{{\mathrm{s}}_i}}} =  {{{{\left| {{g_{{i}}}} \right|}^2}}}/{{{\sigma ^2}}}$ is the interference CNR of SU to PU node $i$ and ${\upsilon _{{{\mathrm{p}}_i}}} = {{{{\left| {{h_{ii}}} \right|}^2}}}/{{{\sigma ^2}}}$ in the RSI CNR of PU node $i$. After evaluating  ${\mathcal C}{_{{{y}_i}}}$ and ${{\mathcal C}_{{{\mathcal{I}}_i}}}$, ${R_{{{\mathrm{p}}_i}}}\left( {{p_{\mathrm{s}}},{\mathcal{C}_x}} \right)$ can be simplified as 
\begin{align}
& {R_{{{\mathrm{p}}_i}}}\left( {{p_{\mathrm{s}}},{\mathcal{C}_x}} \right) = \frac{1}{2} \times \nonumber \\ & {\log _2}\left( {\frac{{{{\left( {{p_i}{\gamma _{{{\mathrm{p}}_i}}} + {p_j}{\upsilon _{{{\mathrm{p}}_j}}} + {p_{\mathrm{s}}}{{\mathcal I}_{{{\mathrm{s}}_j}}} + 1} \right)}^2} - {{\left( {{p_{\mathrm{s}}}{{\mathcal I}_{{{\mathrm{s}}_j}}}{\mathcal{C}_x}} \right)}^2}}}{{{{\left( {{p_j}{\upsilon _{{{\mathrm{p}}_j}}} + {p_{\mathrm{s}}}{{\mathcal I}_{{{\mathrm{s}}_j}}} + 1} \right)}^2} - {{\left( {{p_{\mathrm{s}}}{{\mathcal I}_{{{\mathrm{s}}_j}}}{\mathcal{C}_x}} \right)}^2}}}} \right).
\end{align}
%

Similarly, the SU achievable rate can be expressed as
\begin{align}\label{su_rate}
& {R_{\mathrm{s}}}\left( {{p_{\mathrm{s}}},{\mathcal{C}_x}} \right) = \frac{1}{2}{\log _2}\left( {\frac{{p_{\mathrm{s}}^2{\gamma _{\mathrm{s}}}^2\left( {1 - {\mathcal{C}_x^2}} \right)}}{{{{\left( {{p_1}{{\mathcal I}_{{{\mathrm{p}}_1}}} + {p_2}{{\mathcal I}_{{{\mathrm{p}}_2}}} + 1} \right)}^2}}} } \right. \nonumber \\ & \qquad \qquad \qquad \qquad \qquad 
\left. {+ \frac{{2{p_{\mathrm{s}}}{\gamma _{\mathrm{s}}}}}{{{p_1}{{\mathcal I}_{{{\mathrm{p}}_1}}} + {p_2}{{\mathcal I}_{{{\mathrm{p}}_2}}} + 1}} + 1} \right),
\end{align}
where ${\gamma _{\mathrm{s}}} = {{{{\left| {{g}} \right|}^2}}}/{{{\sigma ^2}}}$ is the SU direct CNR between the SU transmitter and receiver and ${{\mathcal I}_{{{\mathrm{p}}_i}}} = {{{{\left| {{h_{i}}} \right|}^2}}}/{{{\sigma ^2}}}$ is the PU node $i$ interference CNR to the SU.

The direct, interference and RSI CNR ${\gamma _{{{\mathrm{p}}_{_i}}}}, \, { \gamma _{\mathrm{s}}}, \;{ {\mathcal I}_{{{\mathrm{p}}_i}}}, \; { {\mathcal I}_{{{\mathrm{s}}_i}}}, \, { \upsilon _{{{\mathrm{p}}_i}}}$  are  exponentially distributed random variables with mean values ${\bar \gamma _{{{\mathrm{p}}_{_i}}}}, \, {\bar \gamma _{\mathrm{s}}}, \, {\bar {\mathcal I}_{{{\mathrm{p}}_i}}}, \, {\bar {\mathcal I}_{{{\mathrm{s}}_i}}}, \, {\bar \upsilon _{{{\mathrm{p}}_i}}}$ respectively.

One can notice from (\ref{pu_rate}) and (\ref{su_rate}) that if $\mathcal{C}_x=0$, we can obtain the well known formulations of the achievable rates of proper signaling. Moreover, if $\mathcal{C}_x$ increases, the SU rate decreases while the PU rate increases which will allow the SU to increase its transmitted power while satisfying the PU QoS requirements.
\section{Outage probability analysis}\label{out_prob_der}
In this section, we first derive the exact outage probability for the SU, then we derive an upper bound for the~PU. 
\subsection{SU Outage Probability}
Assume that the SU target rate is ${{R_{0,{\mathrm{s}}}}}$, then its outage probability, ${P_{{\mathrm{out,s}}}}\left( {{p_{\mathrm{s}}},{{\mathcal C}_x}} \right)$, is defined as
\begin{align}\label{su_out}
{P_{{\mathrm{out,s}}}}\left( {{p_{\mathrm{s}}},{{\mathcal C}_x}} \right) = \Pr \left\{ {{R_{\mathrm{s}}}\left( {{p_{\mathrm{s}}},{{\mathcal C}_x}} \right) < {R_{0,{\mathrm{s}}}}} \right\}.
\end{align}
By substituting (\ref{su_rate}) in (\ref{su_out}), we obtain
\begin{align}\label{su_out_ineq} 
& {P_{{\mathrm{out,s}}}}\left( {{p_{\mathrm{s}}},{{\mathcal C}_x}} \right) =  \Pr \Bigg\{ \frac{{p_{\mathrm{s}}^2{\gamma _{\mathrm{s}}}^2\left( {1 - {\mathcal{C}_x^2}} \right)}}{{{{\left( {{p_1}{{\mathcal I}_{{{\mathrm{p}}_1}}} + {p_2}{{\mathcal I}_{{{\mathrm{p}}_2}}} + 1} \right)}^2}}}  \nonumber \\
& \qquad \qquad \qquad  \qquad + \frac{{2{p_{\mathrm{s}}}{\gamma _{\mathrm{s}}}}}{{{p_1}{{\mathcal I}_{{{\mathrm{p}}_1}}} + {p_2}{{\mathcal I}_{{{\mathrm{p}}_2}}} + 1}} - {\Gamma _{\mathrm{s}}} < 0 \Bigg\},
\end{align}
where ${\Gamma _{\mathrm{s}}} = {2^{2{R_{0,{\mathrm{s}}}}}} - 1$. One can show that the conditional SU outage probability (conditioned on ${{\mathcal I}_{{{\mathrm{p}}_1}}}$and ${{\mathcal I}_{{{\mathrm{p}}_2}}}$) is given by
\begin{align}\label{su_out_int}
{P_{{\mathrm{out,s}}}}\left( {{p_{\mathrm{s}}},{{\mathcal C}_x}\left| {{{\mathcal I}_{{{\mathrm{p}}_1}}},{{\mathcal I}_{{{\mathrm{p}}_2}}}} \right.} \right) = \int\limits_0^{{\gamma _{{{\mathrm{s}}_0}}}} {\frac{1}{{{{\bar \gamma }_{\mathrm{s}}}}}} \exp \left( { - \frac{x}{{{{\bar \gamma }_{\mathrm{s}}}}}} \right)dx,
\end{align}
where ${\gamma _{{{\mathrm{s}}_0}}}$ is the non-negative zero obtained by solving the quadratic inequality in (\ref{su_out_ineq}) with respect to ${\gamma _{\mathrm{s}}}$, which is found to be
\begin{equation}
{\gamma _{{{\mathrm{s}}_0}}} = \frac{{{p_1}{{\mathcal I}_{{{\mathrm{p}}_1}}} + {p_2}{{\mathcal I}_{{{\mathrm{p}}_2}}} + 1}}{{{p_{\mathrm{s}}}\left( {1 - {{{\mathcal C}_x}^2}} \right)}}\left( {\sqrt {1 + \left( {1 - {{\mathcal{C}_x^2}}} \right){\Gamma _{\mathrm{s}}}}  - 1} \right).
\end{equation}
By evaluating the integral in (\ref{su_out_int}), we obtain
\begin{align}
&{P_{{\mathrm{out,s}}}}\left( {{p_{\mathrm{s}}},{{\mathcal C}_x}\left| {{{\mathcal I}_{{{\mathrm{p}}_1}}},{{\mathcal I}_{{{\mathrm{p}}_2}}}} \right.} \right) = \nonumber \\ & \qquad  \quad 1 - \exp \left( { - \frac{{{p_1}{{\mathcal I}_{{{\mathrm{p}}_1}}} + {p_2}{{\mathcal I}_{{{\mathrm{p}}_2}}} + 1}}{{\left( {1 - {\mathcal{C}_x^2}} \right)}}{\Psi _{\mathrm{s}}}\left( {{p_{\mathrm{s}}},{{\mathcal C}_x}} \right)} \right),
\end{align}
where ${\Psi _{\mathrm{s}}}\left( {{p_{\mathrm{s}}},{{\mathcal C}_x}} \right)$ is defined as
\begin{equation}
{\Psi _{\mathrm{s}}}\left( {{p_{\mathrm{s}}},{{\mathcal C}_x}} \right) =\frac{ {\sqrt {1 + \left( {1 - {\mathcal{C}_x^2}} \right){\Gamma _{\mathrm{s}}}}  - 1}} { {{p_{\mathrm{s}}}{{\bar \gamma }_{\mathrm{s}}}} }.
\end{equation}
By averaging over the exponential statistics of ${{\mathcal I}_{{{\mathrm{p}}_i}}}$, we get
\begin{equation} \label{p_out_s}
{P_{{\mathrm{out,s}}}}\left( {{p_{\mathrm{s}}},{{\mathcal C}_x}} \right) = 1 - {\frac{{{{\left( {1 - {\mathcal {C}_x^2}} \right)}^2} \exp \left( { - \frac{{{\Psi _{\mathrm{s}}}\left( {{p_{\mathrm{s}}},{{\mathcal C}_x}} \right)}}{{  {1 - {\mathcal {C}_x^2}}  }}} \right) }}{{  \prod\limits_{j = 1}^2 {\left( {{p_j}{{\bar {\mathcal I} }_{{{\mathrm{p}}_j}}}{\Psi _{\mathrm{s}}}\left( {{p_{\mathrm{s}}},{{\mathcal C}_x}} \right) + 1 - \mathcal{C}_x^2} \right)}   }}}.
\end{equation}
\subsection{PU Outage Probability}
The outage probability of PU node $i$ for a given target rate $R_{0,{{\mathrm{p}}_i}}$ is defined as
\begin{align}\label{pu_out}
{P_{{\mathrm{out}},{{\mathrm{p}}_i}}}\left( {{p_{\mathrm{s}}},\mathcal{C}_x} \right) = \Pr \left\{ {{R_{{{\mathrm{p}}_i}}}\left( {{p_{\mathrm{s}}},\mathcal{C}_x} \right) < {R_{0,{{\mathrm{p}}_i}}}} \right\}.
\end{align}
Similar to the above subsection, one can show that the outage probability of PU node $i$ (conditioned on ${{\mathcal I}_{{{\mathrm{s}}_j}}}$ and ${\upsilon _{{{\mathrm{p}}_j}}}$) is 
\begin{align} \label{pu_out_cond}
&{P_{{\mathrm{out,}}{{\mathrm{p}}_i}}}\left( {{p_{\mathrm{s}}},{{\mathcal C}_x}\left| {{{\mathcal I}_{{{\mathrm{s}}_j}}},{\upsilon _{{{\mathrm{p}}_j}}}} \right.} \right) = 1 - \nonumber \\  &\exp \left(
{ - \left( {{p_{_j}}{\upsilon _{{{\mathrm{p}}_j}}} + {p_{\mathrm{s}}}{{\mathcal I}_{{{\mathrm{s}}_j}}} + 1} \right){\Psi _{{{\mathrm{p}}_i}}}\left( {\frac{{{p_{\mathrm{s}}}{{\mathcal I}_{{{\mathrm{s}}_j}}}{{\mathcal C}_x}}}{{ {{p_{_j}}{\upsilon _{{{\mathrm{p}}_j}}} + {p_{\mathrm{s}}}{{\mathcal I}_{{{\mathrm{s}}_j}}} + 1} }}} \right)} \right), 
\end{align} 
where ${\Psi _{{{\mathrm{p}}_i}}}\left( x \right) = \left( {\sqrt {1 + {\Gamma _{{\mathrm{p}_i}}}\left( {1 - {x^2}} \right)}  - 1} \right)/\left( {{p_{_i}}{{\bar \gamma }_{{p_i}}}} \right)$ and ${\Gamma _{{\mathrm{p}_i}}} = {2^{2{R_{0,{\mathrm{p}_i}}}}} - 1$. Then, the PU outage probability is found by evaluating the following expression  
\begin{equation}  \label{pu_integral}
{P_{{\mathrm{out,}}{{\mathrm{p}}_i}}}\left( {{p_{\mathrm{s}}},{{\mathcal C}_x}} \right) =  \mathbb{E}_{{{\mathcal I}_{{\mathrm{s}_j}}},{\upsilon _{{{\mathrm{p}}_j}}}}\left\{ {P_{{\mathrm{out,}}{{\mathrm{p}}_i}}}\left( {{p_{\mathrm{s}}},{{\mathcal C}_x}\left| {{{\mathcal I}_{{{\mathrm{s}}_j}}},{\upsilon _{{{\mathrm{p}}_j}}}} \right.} \right)  \right\}
\end{equation}
Unfortunately, there is no closed form expression for this expectation except at $\mathcal{C}_x=0$, which yields
\begin{align} \label{pu_outc0}
{P_{{\mathrm{out,}}{{\mathrm{p}}_i}}}\left( {{p_{\mathrm{s}}},0} \right) = 1 -   {\frac{{\exp \left( { - {\Psi _{{{\mathrm{p}}_i}}}\left( 0 \right)} \right)}}{{\left( {{p_{\mathrm{s}}}{{\bar {\mathcal I}}_{{{\mathrm{s}}_j}}}{\Psi _{{{\mathrm{p}}_i}}}\left( 0 \right) + 1} \right)\left( {{p_{_j}}{{\bar \upsilon }_{{{\mathrm{p}}_j}}}{\Psi _{{{\mathrm{p}}_i}}}\left( 0 \right) + 1} \right)}}}.
\end{align}
For $\mathcal{C}_x\neq 0$, we resort to obtaining an upper bound for the outage probability. First, one can prove the convexity of the exponential term in \eqref{pu_out_cond} in terms of ${{\mathcal I}_{{{\mathrm{s}}_j}}}$ and hence we use the Jensen's inequality to obtain an upper bound of the PU outage probability conditioned on $\upsilon _{{{\rm{p}}_j}}$. Then, we obtain a similar convex exponential term with respect to $\upsilon _{{{\rm{p}}_j}}$, which motivates us to apply the Jensen's inequality again obtaining the following outage probability upper bound  

\begin{align} \label{pu_outage_ub}
& P_{{\mathrm{out,}}{{\mathrm{p}}_i}}^{\mathrm{UB}}\left( {{p_{\mathrm{s}}},{{\mathcal C}_x}} \right) = 
 1 - \nonumber \\ & \exp \left( - \left( {{p_{_j}}{\bar{\upsilon} _{{{\mathrm{p}}_j}}} + {p_{\mathrm{s}}}{{\bar{\mathcal I}}_{{{\mathrm{s}}_j}}} + 1} \right){\Psi _{{{\mathrm{p}}_i}}}\left( {\frac{{{p_{\mathrm{s}}}{{\bar{\mathcal I}}_{{{\mathrm{s}}_j}}}{{\mathcal C}_x}}}{{\left( {{p_{_j}}{\bar{\upsilon} _{{{\mathrm{p}}_j}}} + {p_{\mathrm{s}}}{\bar{{\mathcal I}}_{{{\mathrm{s}}_j}}} + 1} \right)}}} \right) \right).
\end{align}

\section{SU Transmitted Signal Design}
In this section, we optimize the SU signal parameters to minimize the SU outage probability while maintaining a predetermined PU outage probability for each PU link. 

First, we state the PU design criterion in order to be satisfied by the SU during the operation with either proper or improper Gaussian signaling. Assume the PU nodes operate with a rate ${R_{0,{{{\rm{p}}}_i}}}$ and a target maximum outage probability of  ${{\cal O}_{{{{\rm{p}}}_i}}}$ considering an allowable interference power margin ${{\cal P}_{{\rm{int,}}{{{\rm{p}}}_i}}}$. The PU is assumed to use its maximum budget $p_i$ to guarantee achieving its required QoS. Thus, the target maximum PU outage probability,  from its perspective,  is expressed as  
\begin{align} \label{pu_out_margin}
{{\cal O}_{{{{\rm{p}}}_i}}} &= \Pr \left\{ {{{\log }_2}\left( {1 + \frac{{{p_i}{{\left| {{h_{ij}}} \right|}^2}}}{{{\sigma ^2} + {{\cal P}_{{\rm{int,}}{{{\rm{p}}}_i}}}}}} \right) < {R_{0,{{{\rm{p}}}_i}}}} \right\} \nonumber \\
&= 1 - \exp \Bigg( { -  \frac{{1 + {{\cal I}_{\max ,{{{\rm{p}}}_i}}}}}{{{{\bar \gamma }_{{{\rm{p}}}_i}}}} {\Psi _{{{{\rm{p}}}_i}}}\left( 0 \right)} \Bigg),
\end{align}
where ${{\cal I}_{\max ,{{{\rm{p}}}_i}}}={{\cal P}_{{\rm{int,}}{{{\rm{p}}}_i}}}/{\sigma ^2}$ is the maximum allowable margin interference-to-noise ratio at the receiver of PU node $i$. By considering a maximum PU outage probability threshold ${{\cal O}_{{{{\rm{p}}}_i}}}$, ${{\cal I}_{\max ,{{{\rm{p}}}_i}}}$ can be found from \eqref{pu_out_margin} as
\begin{align}
{{\cal I}_{\max ,{{{\rm{p}}}_i}}} = \left[\frac{{{\mu _i}}}{{\sqrt {1 + {\Gamma _{{{\rm{p}}}_i}}}  - 1}} - 1\right]^+,
\end{align}
where $[z]^+=\max(0,z)$, ${\mu _i} = {p_i}{{\bar \gamma }_{{{\rm{p}}}_i}}\log \left( {\frac{1}{{1 - {{\cal O}_{{p_i}}}}}} \right)$ and $\log\left(.\right)$ is the natural logarithm. In the following subsections, we design proper and improper Gaussian signals for the SU to improve its performance,  measured by the outage probability, considering a predetermined ${{\cal O}_{{{{\rm{p}}}_i}}}$ and other system parameters such as $p_i$ and ${R_{0,{{{\rm{p}}}_i}}}$.
\subsection{Proper Gaussian Signaling Design}
In the case of proper Gaussian signaling, the SU allocates its power in order to minimize its outage probability subject to its own power budget ${p_{{\rm{s,max}}}}$ and PU QoS by solving the following  optimization problem,
\begin{align} \label{opt_prob_proper}
& \mathop {\min }\limits_{{p_{\rm{s}}}} \quad {P_\mathrm{out,s}}\left( {{p_{\rm{s}}}, 0} \right) \hspace{0.25cm} \nonumber \\
& \mathrm{s. \; t.} \quad {P_{\mathrm{out},{{\mathrm{p}}_i}}}\left( {{p_{\mathrm{s}}},0} \right) \le {{\cal O}_{{\mathrm{p}_i}}},\quad 0 < {p_{\mathrm{s}}} \le {p_{{\mathrm{s,max}}}}.
\end{align}
The predetermined PU outage probability  constraints in \eqref{opt_prob_proper} reduce to ${p_{\rm{s}}}\leq{p^{\left(i\right)}_{\rm{s}}}$, where ${p^{\left(i\right)}_{\rm{s}}}$ is the maximum allowable power that satisfies the PU required  outage probability threshold, which is expressed as 
\begin{align} \label{proper_psi}
{p^{\left(i\right)}_{\rm{s}}}=\left[ \frac{{\exp \left( { - {\Psi _{{{\rm{p}}_i}}}\left( 0 \right)} \right) - {\left( {1 - {{\cal O}_{{{\rm{p}}_i}}}} \right)\left( {{p_{_j}}{{\bar \upsilon }_{{{\rm{p}}_j}}}{\Psi _{{{\rm{p}}_i}}}\left( 0 \right) + 1} \right)} }}{{{{\bar {\cal I}}_{{{\rm{s}}_j}}}{\Psi _{{{\rm{p}}_i}}}\left( 0 \right)\left( {1 - {{\cal O}_{{{\rm{p}}_i}}}} \right)\left( {{p_{_j}}{{\bar \upsilon }_{{{\rm{p}}_j}}}{\Psi _{{{\rm{p}}_i}}}\left( 0 \right) + 1} \right)}} \right]^+.
\end{align}
Thus, we can rewrite the optimization problem in \eqref{opt_prob_proper} as  
\begin{align} \label{opt_prob_proper_equivalent}
\mathop {\min }\limits_{{p_{\rm{s}}}} \; {P_\mathrm{out,s}}\left( {{p_{\rm{s}}}, 0} \right)  \; \;  \mathrm{s. \;  t. \;} \; {p_{\rm{s}}} \leq \mathop {\min } \left( p_{\rm{s}}^{\left( 1 \right)}, p_{\rm{s}}^{\left( 2 \right)}, p_{\mathrm{s,max}} \right).
\end{align}
One can prove that ${P_\mathrm{out,s}}\left( {{p_{\rm{s}}}, 0} \right)$ is strictly decreasing in $p_\mathrm{s}$, thus the upper bound of the constraint achieves the optimal minimum SU outage probability and is expressed as 
\begin{align} 
{p_{\rm{s}}} = \mathop {\min } \left( p_{\rm{s}}^{\left( 1 \right)}, p_{\rm{s}}^{\left( 2 \right)}, p_{\mathrm{s,max}} \right).
\end{align}
From (\ref{proper_psi}), the SU operates if $\exp \left( { - {\Psi _{{{\rm{p}}_i}}}\left( 0 \right)} \right) > \left( {\left( {1 - {{\cal O}_{{{\rm{p}}_i}}}} \right)\left( {{p_{_j}}{{\bar \upsilon }_{{{\rm{p}}_j}}}{\Psi _{{{\rm{p}}_i}}}\left( 0 \right) + 1} \right)} \right)$. Otherwise, it remains~
silent. 
\subsection{Improper Gaussian Signaling Design}
The improper Gaussian signal design aims to tune  $p_{\mathrm{s}}$ and $\mathcal{C}_x$ to minimize the SU outage while holding a required PU QoS based on the upper bound derived  in \eqref{pu_outage_ub}  achieving the worst case system design. To this end, we formulate the design optimization problem as,
\begin{align} \label{opt_prob}
&\mathop {\min }\limits_{{p_{\rm{s}}},{{\cal C}_x}} {P_\mathrm{out,s}}\left( {{p_{\rm{s}}},{{\cal C}_x}} \right) \hspace{0.25cm} \nonumber \\
&\mathrm{s. \; t. \;}{P_{\mathrm{out},{{\rm{p}}_i}}^{\rm{UB}}}\left( {{p_{\rm{s}}},\;{\cal C}_x} \right) \le {{\cal O}_{{\mathrm{p}_i}}},\;0 \le {p_{\rm{s}}} \le {p_{{\rm{s,max}}}}, \;0 \leq {{\cal C}_x} \le 1.
\end{align}
Unfortunately, this optimization problem turns to be non-linear and non-convex which makes it hard to find its optimal solution. However, similar to what we did in the proper signaling design, we exploit some monotonicity properties of the objective function and the constraints that lead us to the optimal solution of \eqref{opt_prob}. 

First, based on \eqref{pu_outage_ub}, we express the outage probability constraints in \eqref{opt_prob}  by the following equivalent quadratic inequality in terms of~${p_{\rm{s}}}$ as
\begin{align}\label{ps_improper_quadraric}
{\Gamma _{{\mathrm{p}_i}}}\left( {1 - \mathcal{C}_x^2} \right)\bar {\cal I} _{{{\rm{s}}_j}}^2 p_\mathrm{s}^2 + 2{\Lambda _i} {\bar {\cal I} _{{{\rm{s}}_j}}} p_\mathrm{s} - {\Upsilon _i} \le 0,
\end{align}
where ${\Upsilon _i} = \left( {\mu _i^2 + 2{\beta _j}{\mu _i} - {\Gamma _{{\mathrm{p}_i}}}\beta _j^2} \right)$, ${\Lambda _i} = \left({\beta _j}{\Gamma _{{\mathrm{p}_i}}} - {\mu _i}\right)$, ${\beta _j} = \left({p_j}{{\bar \upsilon }_{{{\rm{p}}_j}}} + 1\right)$. Based on \eqref{ps_improper_quadraric}, the outage probability constraints is equivalent to ${p_{\rm{s}}}\leq{p^{ (i )}_{\rm{s}}}\left({{\cal C}_x}\right)$, where ${p_{\rm{s}}^{(i)}}\left({{\cal C}_x}\right)$ is found by equating the left-hand-side of \eqref{ps_improper_quadraric} to zero and find the feasible root(s). One can show that if $\Upsilon _i < 0$, then, $ {\Lambda _i}>0$, which results in two negative roots. On the other hand, if $\Upsilon _i > 0$, then, there is exactly one positive and one negative roots. Thus, the power upper bound exists at  $\Upsilon _i > 0$ and is given by   
\begin{align}\label{ps_improper_max}
p_\mathrm{s}^{\left( i \right)}\left({{\cal C}_x}\right) =  \frac{{\sqrt {\Lambda _i^2 + {\Gamma _{{\mathrm{p}_i}}}\left( {1 - {\cal C}_x^2} \right){\Upsilon _i}}  - {\Lambda _i}}}{{{\Gamma _{{\mathrm{p}_i}}}{{\bar {\cal I}}_{{{\rm{s}}_j}}}\left( {1 - {\cal C}_x^2} \right)}}.
\end{align}
Thus, the power constraints in \eqref{opt_prob} is equivalently rewritten~as 
\begin{align}\label{ps_min}
{p_{\rm{s}}}\left({{\cal C}_x}\right) \le \min \left\{ {p_{\rm{s}}^{\left( 1 \right)}\left({{\cal C}_x}\right),\;p_{\rm{s}}^{\left( 2 \right)}}\left({{\cal C}_x}\right),\;{p_{{\rm{s,max}}}} \right\}. 
\end{align}
 We obtain the distinct intersection points of these three functions in $0 < {{\cal C}_x} < 1$. First, we can show that $p_{\rm{s}}^{\left( i \right)}$ is a strictly increasing function in ${\cal C}_x$. Hence, \eqref{ps_min} can be described as a piecewise function with a maximum of four intervals (three breaking points) and a minimum of one interval (zero breaking points). The intersection point, $r^{\left( i \right)}$, between $p_{\rm{s}}^{\left( i \right)}$ and ${p_{{\rm{s,max}}}}$~is 
\begin{align}
r^{\left( i \right)} = \sqrt {1 + \frac{{2\left( {{p_{{\rm{s,max}}}}{{\bar {\cal I}}_{{{\rm{s}}_j}}}} \right){\Lambda _i} - {\Upsilon _i}}}{{{\Gamma _{{\mathrm{p}_i}}}{{\left( {{p_{{\rm{s,max}}}}{{\bar {\cal I}}_{{{\rm{s}}_j}}}} \right)}^2}}}},  
\end{align}   
which exists  if $p_s^{\left( i \right)}\left(0\right) < {p_{{\rm{s,max}}}}$ and $p_\mathrm{s}^{\left( i \right)}\left(1\right) > {p_{{\rm{s,max}}}}$.
Furthermore, the intersection between $p_{\rm{s}}^{\left( 1 \right)}$ and $p_{\rm{s}}^{\left( 2 \right)}$ in the interested interval, if they are not identical, is $r^{\left( 3 \right)} = \sqrt{(1 - \kappa)}$, where $\kappa$ is computed from

\footnotesize
\begin{align} \label{ps_i_intersection}
\kappa = \frac{{4{\Gamma _{{\mathrm{p}_1}}}{\Gamma _{{\mathrm{p}_2}}}{{\bar {\cal I}}_{{{\rm{s}}_1}}}{{\bar {\cal I}}_{{{\rm{s}}_2}}}\left( {{\Gamma _{{\mathrm{p}_1}}}{\Lambda _2}{{\bar {\cal I}}_{{{\rm{s}}_2}}} - {\Gamma _{{\mathrm{p}_2}}}{\Lambda _1}{{\bar {\cal I}}_{{{\rm{s}}_1}}}} \right)\left( {{\Lambda _2}{\Upsilon _1}{{\bar {\cal I}}_{{{\rm{s}}_1}}} - {\Lambda _1}{\Upsilon _2}{{\bar {\cal I}}_{{{\rm{s}}_2}}}} \right)  }}{{{{\left( {{\Gamma _{{\mathrm{p}_2}}}{\Upsilon _1}\bar {\cal I}_{{{\rm{s}}_1}}^2 - {\Gamma _{{\mathrm{p}_1}}}{\Upsilon _2}\bar {\cal I}_{{{\rm{s}}_2}}^2} \right)}^2}}}
\end{align} \normalsize
which exists if $p_s^{\left( i \right)}\left( 0 \right) < p_s^{\left( j \right)}\left( 0 \right)$ and $p_s^{\left( i  \right)}\left( 1 \right) > p_s^{\left( j \right)}\left( 1 \right)$.

Define the interval boundaries   points as ${\cal C}_ x^{(z)}$, where $z$ is an integer number in  $ [1,k+1]$, $k$ is the number of  distinct  intersection points, i.e. $k \in \{ 0,1,2,3 \}$, ${\cal C}_ x ^{(0)}=0$, ${\cal C}_x ^{(k+1)}=1$ and ${\cal C}_ x^{(1)}$, ${\cal C}_ x^{(2)}$ and ${\cal C}_ x^{(3)}$ are the ordered  distinct  intersection points (if~exist). 

Thereafter, we divide the optimization problem in  \eqref{opt_prob} into $(k+1)$ subproblems, where  each subproblem is defined in a specific range ${\cal C}_x ^{(z-1)} \le {{\cal C}_x} \le {\cal C}_x ^{(z )}$. We can show that ${P_\mathrm{out,s}}\left( p_{\mathrm{s}}, {{{\cal C}_x}} \right)$ is strictly decreasing in $p_{\mathrm{s}}$ for a fixed ${{{\cal C}_x}}$. Hence, $p_\mathrm{s}$ is assigned the upper bound of \eqref{ps_min}  to minimize the outage probability obtaining $k+1$ subproblmes, where the $z^{\mathrm{th}}$ subproblem is written as
\begin{align} \label{opt_subprob}
\mathfrak{P}_{z}:\mathop {\min }\limits_{{{\cal C}_x}} {P_\mathrm{out,s}}\left( {{{\cal C}_x}} \right) \; \mathrm{s. \; t. \; } \; {\cal C}_x^{(z-1)} \le {{\cal C}_x} \le {\cal C}_ x ^{(z)}.
\end{align}
To solve the $\mathfrak{P}_{z}$ problem, we have two cases, either $p_{\rm{s}}={p_{{\rm{s,max}}}}$ or $p_{\rm{s}}=p_{\rm{s}}^{\left( i \right)}\left({{\cal C}_x}\right)$. If $p_{\rm{s}}= p_{\rm{s}}^{\left( i \right)}$ in  \eqref{ps_min}, ${P_\mathrm{out,s}}\left( {{{\cal C}_x}}\right)$ is a strictly decreasing function in $\mathcal{C}_x$ if\footnote{See Appendix A for the proof.}
\begin{align} \label{improper_condition}
\Lambda _i < 0 \hspace{0.25cm} \mathrm{or} \hspace{0.25cm} {\Gamma _{\rm{s}}} < \frac{{{\Gamma _{{\mathrm{p}_i}}}{\Upsilon _i}}}{\Lambda _i^2}.
\end{align} 
Therefore,  the optimal solution pair in this case is $ ( {p_{\rm{o}}^{(z)},{\cal C}_{\rm{o}}^{(z)}} )= (p_s^{\left( i \right)} ({\cal C}_ x ^{(z)} ), {\cal C}_x ^{(z)} )$.  Otherwise, it is a strictly increasing function and hence,  $ ( {p_{\rm{o}}^{(z)},{\cal C}_{\rm{o}}^{(z)}} )= (p_s^{\left( i \right)} ({\cal C}_ x ^{(z-1)} ), {\cal C}_x ^{(z-1)} )$. Moreover, one can show easily in a similar way to the proof in Appendix A, that if $p_{\rm{s}}={p_{{\rm{s,max}}}}$ in \eqref{ps_min}, ${P_\mathrm{out,s}}\left( {{{\cal C}_x}}\right)$ is a strictly increasing function in $\mathcal{C}_x$,  hence, the optimal solution pair is $( {p_{\rm{o}}^{(z)},{\cal C}_{\rm{o}}^{(z)}} )= ({p_{{\rm{s,max}}}}, {\cal C}_ x^{(z-1)})$. At the end, we pick the  optimal pair $( {p_{\rm{o}}^{(z)},{\cal C}_{\rm{o}}^{(z)}} )$ that minimizes the objective function ${P_\mathrm{out,s}}\left( {{p_{\rm{s}}},{{\cal C}_x}}\right)$. Based on the aforementioned analysis, we develop Algorithm I to find the optimal solution pairs in $z$ regions, then find the pair, $\left( {p_{\rm{s}}^*,{\cal C}_x^*} \right)$, with minimum SU outage probability. 
 
\floatname{algorithm}{}
\begin{algorithm} \label{alg2}
\renewcommand{\thealgorithm}{}
\newcommand{\tab}[1]{\hspace{.06\textwidth}\rlap{#1}}
\caption{\textbf{Algorithm I}}
\begin{algorithmic}[1]
\State \textbf{Input} $p_i$, ${\bar \gamma _{{{\mathrm{p}}_{_i}}}}$, ${{\bar {\cal I}}_{{{\rm{p}}_i}}}$, ${\bar{ \mathcal I}_{{{\mathrm{s}}_i}}}$, ${\bar \upsilon _{{{\mathrm{p}}_i}}}$, ${R_{0,{{\rm{p}}_i}}}$,  ${{\cal O}_{{\mathrm{p}_i}}}$, ${\cal C}_ x^{(z)}$ , $p_{\mathrm{s}}^{(0)} = {p_{{\rm{s,max}}}}$,   ${\bar \gamma _{\rm{s}}}$, and ${R_{0,{{\rm{s}}}}}$
\For {$z=1:k+1$}
\State \textbf{Construct} interval $ [{\cal C}_ x ^{(z-1)} , {\cal C}_x ^{(z)} ]$
\State \textbf{Compute} $m = \mathop {\arg \min }\limits_{l \in \{0,1,2\}}   p_{\mathrm{s}}^{( l  )} \left(  \frac{{\cal C}_ x^{(z-1)} +{\cal C}_ x^{(z)}}{2} \right)$ 
\If{ $m =0$}
\State $p_{\rm{o}}^{(z)}\leftarrow {p_{{\rm{s,max}}}}$,   $ \quad \mathcal{C}_\mathrm{o}^{(z)}\leftarrow {\cal C}_ x ^{(z-1)}$
\ElsIf{ $\Lambda _m < 0 \hspace{0.25cm} \mathrm{or} \hspace{0.25cm} {\Gamma _{\rm{s}}} < \frac{{{\Gamma _{{\mathrm{p}_m}}}{\Upsilon _m}}}{\Lambda _m^2} $ is true}
\State $p_{\rm{o}}^{(z)}\leftarrow p_\mathrm{s}^{\left( m \right)}\left({\cal C}_x ^{(z)}\right)$, $\quad \mathcal{C}_{\rm{o}}^{(z)}\leftarrow {\cal C}_x ^{(z)}$
\Else {}
\State $p_{\rm{o}}^{(z)}\leftarrow p_{\mathrm{s}}^{\left( m \right)}\left({\cal C}_x ^{(z-1)}\right)$, $\quad\mathcal{C}_{\rm{o}}^{(z)}\leftarrow {\cal C}_x ^{(z-1)}$
\EndIf 
\EndFor
\State \textbf{Output} $\left( {p_{\rm{s}}^*,{\cal C}_x^*} \right) = \mathop {\arg \min }\limits_{p_{\rm{o}}^{(z)},\;{\cal C}_o^{(z)}} {p_{{\rm{out}},{\rm{s}}}}\left( {p_{\rm{o}}^{(z)},{\cal C}_{\rm{o}}^{(z)}} \right)$
  \end{algorithmic}
\end{algorithm}
     
\section{Numerical Results}
In this section, we present some numerical examples which validate the introduced analysis and investigate the benefits of employing improper Gaussian signaling in spectrum sharing with FD PU. Throughout this Section, we use the following general system parameters for all examples, unless otherwise specified. For the PU nodes, we assume ${R_{0,{{\rm{p}}_i}}}=0.5$ b/s/Hz with a maximum power budget $p_i=1 \; W$. The communications channels are characterized as, ${\bar \gamma _{{{\mathrm{p}}_{_i}}}}=25$ dB, ${{\bar {\cal I}}_{{{\rm{p}}_i}}}=3$ dB, ${\bar \upsilon _{{{\mathrm{p}}_i}}}=5$ dB. The SU is assumed to target ${R_{0,{{\rm{s}}}}}=0.5$ b/s/Hz using $p_{\rm{s,max}}=1 \; W$. The SU channels parameters are assumed to be ${{\bar {\cal I}}_{{{\rm{s}}_i}}}=13$ dB and ${\bar \gamma _{{{\mathrm{s}}}}}=20$ dB.

\begin{figure}[!t]
\centering
\includegraphics[width=9cm]{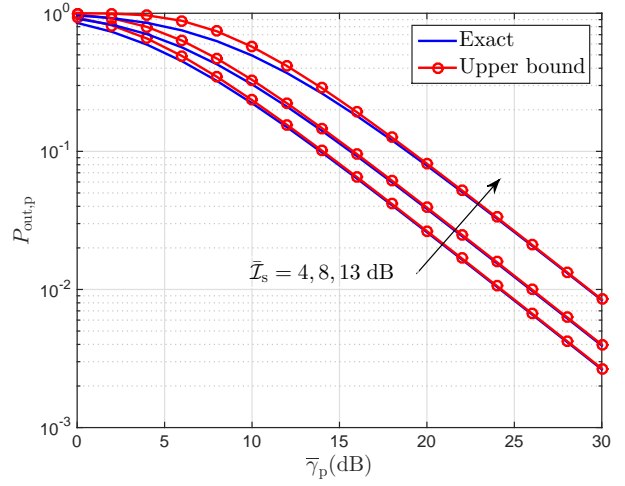}
\caption{A comparison between the exact PU outage probability and the upper bound versus ${\bar \gamma _{{{\mathrm{p}}}}}$ for ${{\bar {\cal I}}_{{{\rm{s}}}}}=4, 8, 13$.}
\label{SimEx1}
\end{figure}

\textit{Example 1:} This example compares the upper bound computed from \eqref{pu_outage_ub} with the exact value computed by evaluating the expectations in \eqref{pu_integral} numerically. We assume ${\bar \gamma _{{{\mathrm{p}}_{_i}}}}={\bar \gamma _{{{\mathrm{p}}}}}$, $\mathcal{C}_x=0.5$ and ${{\bar {\cal I}}_{{{\rm{s}}_i}}}={{\bar {\cal I}}_{{{\rm{s}}}}}=4, 8, 13$ dB. As shown in Fig. \ref{SimEx1}, the upper bound is tight to the exact outage probability for different ${{\bar {\cal I}}_{{{\rm{s}}}}}$. Similar results are observed for different ${R_{0,{{\rm{p}}_i}}}$ but are not included due to space limitations.

\begin{figure}[!t]
\centering
\includegraphics[width=9cm]{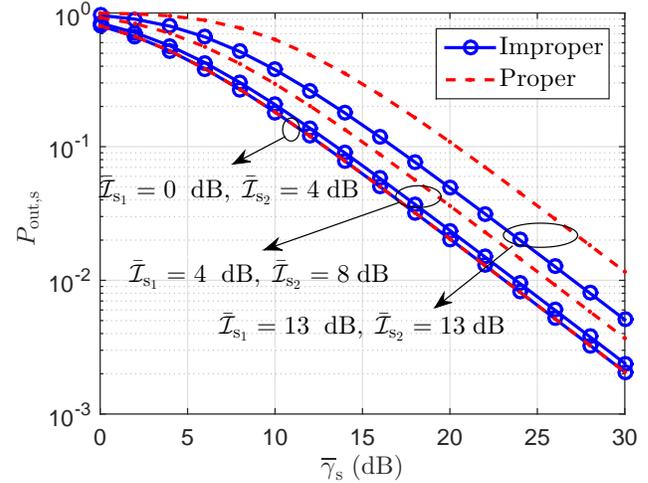}
\caption{SU outage probability for proper and improper Gaussian signaling versus ${\bar \gamma _{{{\mathrm{s}}}}}$ for different pairs of $\bar {{\cal I}}_{{{\rm{s}}_i}}$.}
\label{SimEx2}
\end{figure}

\textit{Example 2:} In this example, we inspect the benefits of designing the improper Gaussian signaling for SU over the conventional proper Gaussian signaling design. First, we assume that the required PU outage probability threshold ${{\cal O}_{{\mathrm{p}_i}}}=0.01$, $\left(\bar {{\cal I}}_{{{\rm{s}}_1}},\bar {{\cal I}}_{{{\rm{s}}_2}}\right)$ are assumed to have $\left(0,4\right), \left(4,8\right), \left(13,13\right)$ dB. The proper design is based on \eqref{proper_psi}. For the improper design, we first obtain the distinct intersection points, if exist, and sort them in $\mathcal{C}_x^{z}$, then we apply Algorithm I to obtain the optimal pair $\left(p_{{\rm{s}}}^*,\mathcal{C}_x^*\right)$. Fig. \ref{SimEx2} shows the SU outage probability versus ${\bar \gamma _{{{\mathrm{s}}}}}$ for different pairs of ${{\bar {\cal I}}_{{{\rm{s}_i}}}}$. For $\left(\bar {{\cal I}}_{{{\rm{s}}_1}},\bar {{\cal I}}_{{{\rm{s}}_2}}\right)=\left(0,4\right)$ dB, there is no gain from using improper signaling. In this case, the interference channel is week, which allows the SU with proper signaling to improve its performance (minimize its outage probability) by increasing the transmitted power, which could reach its maximum budget. As we observed from the improper design, $p_\mathrm{s}$ tends to increase with $\mathcal{C}_x$  as can be seen in \eqref{ps_improper_max}, but since $p_\mathrm{s}(0) \simeq p_{\mathrm{s,max}}$, then the improper solution reduces approximately to the proper design. As the the SU interference channels $\bar {{\cal I}}_{{{\rm{s}}_i}}$ become stronger proper, signaling tends to use less power to meet the PU QoS. On the other hand, improper signaling system uses more power to improve its outage probability performance while compensating for its interference impact on the PU by increasing the circularity coefficient. Fig. \ref{SimEx2} shows a $1.5-3.5$ dB improvement as a result of adopting the improper Gaussian signaling.

\begin{figure}[!t]
\centering
\includegraphics[width=9cm]{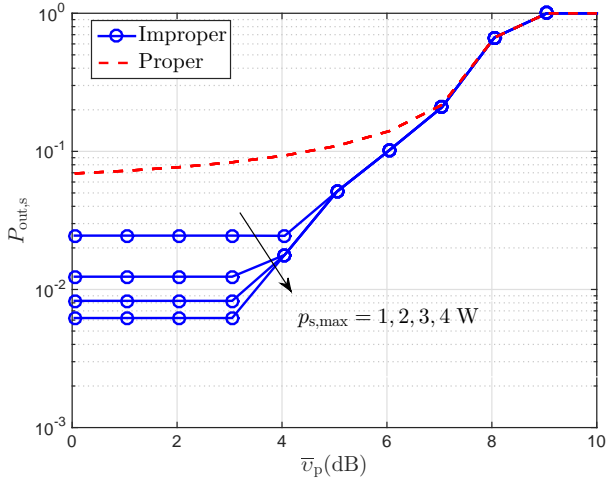}
\caption{SU outage probability for proper and improper Gaussian signaling versus ${\bar \upsilon _{{{\mathrm{p}}}}}$ for different pairs of $p_{\rm{s,max}}$.}
\label{SimEx3}
\end{figure}

\textit{Example 3:} This example investigates the impact of RSI-CNR in limiting the CR operation and compares between its effect on both proper and improper Gaussian signaling based systems. For this purpose, we assume ${\bar \upsilon _{{{\mathrm{p}_i}}}}={\bar \upsilon _{{{\mathrm{p}}}}}$ and plot the SU outage probability versus ${\bar \upsilon _{{{\mathrm{p}}}}}$ for different values of $p_{\rm{s,max}}$ in Fig. \ref{SimEx3}. We observe that improper Gaussian signaling achieves better performance than the proper Gaussian signaling system at low values of ${\bar \upsilon _{{{\mathrm{p}}}}}$. Although the proper Gaussian signaling system cannot get benefits from increasing the power budget, the improper Gaussian signaling tends to use the total budget efficiently and relieve the interference effect on PU by increasing $\mathcal{C}_x$, which compensates for the interference impact as can be seen from \eqref{pu_rate}. On the other hand, at high RSI-CNR values, both proper and improper fail to operate properly.                           
%

\section{Conclusion}
In this paper, we study the opportunity of sharing the spectrum resources of licensed FD PU using improper Gaussian signaling in underlay CR mode. We use the outage probability as a performance metric, then  derive a closed form for the SU and a tight upper bound for the PU. Based on the average CSI, we optimize the SU signal parameters, i.e., transmit power and circularity coefficient, to minimize the SU outage probability while maintaining a predetermined QoS requirements for the PU. As a result, we derive a low complex algorithm that tune the signal parameters to accomplish the design objectives. The numerical results show a promising performance for the improper Gaussian signaling with the average CSI. Specifically, the main advantage of the proposed scheme is for strong SU interference channels to PU, where proper signaling scheme tends to use less transmitted power, while SU with improper Gaussian signaling uses more power and compensate its interference impact through the increase of signal impropriety.

\section*{Appendix A}
In this appendix, we derive the conditions in \eqref{improper_condition} over the interested interval $0 < {{\cal C}_x} < 1$. First, \eqref{p_out_s} can be written as
\begin{align}
{P_{{\rm{out,s}}}}\left( {\mathcal{G},\mathcal{Y}} \right) = 1 - \frac{{{\mathcal{G}^2}{\mathcal{Y}^2}}}{ \prod\limits_{j = 1}^2 {\left( {{p_j}{{\bar {\cal I}}_{{{\rm{p}}_j}}} + \mathcal{G}\mathcal{Y}} \right)} }\exp \left( { - \frac{1}{{\mathcal{G}\mathcal{Y}}}} \right),
\end{align}
where $\mathcal{Y}  = \sqrt {1 - {\cal C}_x^2} /\left( {\sqrt {1 + \left( {1 - {\cal C}_x^2} \right){\Gamma _{\rm{s}}}}  - 1} \right)$ and $\mathcal{G}  = p_{\rm{s}}^{(i)}{{\bar \gamma }_{\rm{s}}}\sqrt {1 - {\cal C}_x^2}$. By using the chain rule of partial derivatives defined as
\begin{align}
\frac{{\partial {P_{{\rm{out,s}}}}}}{{\partial {{\cal C}_x}}} = \frac{{\partial {P_{{\rm{out,s}}}}}}{{\partial \mathcal{G}}}\frac{{d\mathcal{G}}}{{d{{\cal C}_x}}} + \frac{{\partial {P_{{\rm{out,s}}}}}}{{\partial \mathcal{Y}}}\frac{{d\mathcal{Y}}}{{d{{\cal C}_x}}}.
\end{align} 
We obtain,
\begin{align}\label{der_pout_c}
&\frac{{\partial {P_{{\rm{out,s}}}}\left( {\mathcal{G},\mathcal{Y}} \right)}}{{\partial {{\cal C}_x}}} = \left( {\frac{{{  T}{{\cal C}_x}{{\bar \gamma }_{\rm{s}}}\left( { - {\Lambda _i} + \sqrt {\Lambda _i^2 + {\Phi _i}} } \right)}}{{{\Gamma _{{p_j}}}{{\bar {\cal I}}_{{{\rm{s}}_j}}}\left( {\sqrt {1 + \left( {1 - {\cal C}_x^2} \right){\Gamma _{\rm{s}}}}  - 1} \right)\left( {1 - {\cal C}_x^2} \right)}}} \right) \nonumber \\
 & \qquad \qquad  \quad  \times \left( {\frac{{{\Lambda _i}}}{{\sqrt {\Lambda _i^2 + {\Phi _i}} }} - \frac{1}{{\sqrt {1 + \left( {1 - {\cal C}_x^2} \right){\Gamma _{\rm{s}}}} }}} \right),
\end{align}
where
\begin{align}
T = \frac{{\mathcal{G}\mathcal{Y}\left( {\prod\limits_{j = 1}^2 {{p_j}{{\bar {\cal I}}_{{{\rm{p}}_j}}}}  + \Theta  - {\mathcal{G}^2}{\mathcal{Y}^2}} \right) + \Theta }}{{{\Theta ^2}}}\exp \left( { - \frac{1}{{\mathcal{G}\mathcal{Y}}}} \right) > 0, \nonumber
\end{align}
${\Phi _i} = {\Gamma _{{\mathrm{p}_i}}}\left( {1 - {\cal C}_x^2} \right){\Upsilon _i} > 0$ and $\Theta  = \prod\limits_{j = 1}^2 {\left( {{p_j}{{\bar {\cal I}}_{{{\rm{p}}_j}}} + \mathcal{G}\mathcal{Y}} \right)} > 0 $. 

It is clear from \eqref{der_pout_c} that if ${\Lambda _i}<0$, ${P_{{\rm{out,s}}}}$ is directly a strictly decreasing function in $\mathcal{C}_x$ while if ${\Lambda _i}>0$, one can show easily that if ${\Gamma _{\rm{s}}} < \left( {{\Gamma _{{\mathrm{p}_i}}}{\Upsilon _i}} \right)/\Lambda _i^2$, ${P_{{\rm{out,s}}}}$ is also a strictly decreasing function in $\mathcal{C}_x$. Otherwise, It is a strictly increasing function in $\mathcal{C}_x$ and this concludes the proof.

\bibliographystyle{IEEEtran}

\bibliography{IEEEabrv,mgaafar_April_2015}

\begin{thebibliography}{10}
\providecommand{\url}[1]{#1}
\csname url@samestyle\endcsname
\providecommand{\newblock}{\relax}
\providecommand{\bibinfo}[2]{#2}
\providecommand{\BIBentrySTDinterwordspacing}{\spaceskip=0pt\relax}
\providecommand{\BIBentryALTinterwordstretchfactor}{4}
\providecommand{\BIBentryALTinterwordspacing}{\spaceskip=\fontdimen2\font plus
\BIBentryALTinterwordstretchfactor\fontdimen3\font minus
  \fontdimen4\font\relax}
\providecommand{\BIBforeignlanguage}[2]{{%
\expandafter\ifx\csname l@#1\endcsname\relax
\typeout{** WARNING: IEEEtran.bst: No hyphenation pattern has been}%
\typeout{** loaded for the language `#1'. Using the pattern for}%
\typeout{** the default language instead.}%
\else
\language=\csname l@#1\endcsname
\fi
#2}}
\providecommand{\BIBdecl}{\relax}
\BIBdecl

\bibitem{zhao2007survey}
Q.~Zhao and B.~M. Sadler, ``A survey of dynamic spectrum access,'' \emph{{IEEE}
  Signal Process. Mag.}, vol.~24, no.~3, pp. 79--89, May 2007.

\bibitem{kim2012optimal}
H.~Kim, S.~Lim, H.~Wang, and D.~Hong, ``Optimal power allocation and outage
  analysis for cognitive full duplex relay systems,'' \emph{{IEEE} Trans.
  Wireless Commun.}, vol.~11, no.~10, pp. 3754--3765, Oct. 2012.

\bibitem{zhongTOAPPEARperformance}
B.~Zhong, Z.~Zhang, X.~Chai, Z.~Pan, K.~Long, and H.~Cao, ``Performance
  analysis for opportunistic full-duplex relay selection in underlay cognitive
  networks,'' \textit{IEEE Trans. Veh. Technol.} DOI: 10.1109/TVT.2014.2368584.

\bibitem{sabharwal2014band}
A.~Sabharwal, P.~Schniter, D.~Guo, D.~W. Bliss, S.~Rangarajan, and R.~Wichman,
  ``In-band full-duplex wireless: Challenges and opportunities,'' \emph{{IEEE}
  J. Sel. Areas Commun.}, vol.~32, no.~9, pp. 1637--1652, Sep. 2014.

\bibitem{afifi2015incorporating}
W.~Afifi and M.~Krunz, ``Incorporating self-interference suppression for
  full-duplex operation in opportunistic spectrum access systems,''
  \emph{{IEEE} Trans. Wireless Commun.}, vol.~14, no.~4, pp. 2180--2191, April
  2015.

\bibitem{zeng2013transmit}
Y.~Zeng, C.~M. Yetis, E.~Gunawan, Y.~L. Guan, and R.~Zhang, ``{Transmit
  optimization with improper Gaussian signaling for interference channels},''
  \emph{{IEEE} Trans. Signal Process.}, vol.~61, no.~11, pp. 2899--2913, Jun.
  2013.

\bibitem{ho2012improper}
Z.~Ho and E.~Jorswieck, ``{Improper Gaussian signaling on the two-user SISO
  interference channel},'' \emph{{IEEE} Trans. Wireless Commun.}, vol.~11,
  no.~9, pp. 3194--3203, Sep. 2012.

\bibitem{zeng2013optimized}
Y.~Zeng, R.~Zhang, E.~Gunawan, and Y.~Guan, ``{Optimized transmission with
  improper Gaussian signaling in the K-user MISO interference channel},''
  \emph{{IEEE} Trans. Wireless Commun.}, vol.~12, no.~12, pp. 6303--6313, Dec.
  2013.

\bibitem{lameiro2015benefits}
C.~Lameiro, I.~Santamaria, and P.~Schreier, ``Benefits of improper signaling
  for underlay cognitive radio,'' \emph{IEE Wireless Commun. Lett.}, vol.~4,
  no.~1, pp. 22--25, Feb. 2015.

\bibitem{Neeser1993proper}
F.~D. Neeser and J.~L. Massey, ``Proper complex random processes with
  applications to information theory,'' \emph{{IEEE} Trans. Inf. Theory},
  vol.~39, no.~4, pp. 1293--1302, Jul. 1993.

\bibitem{day2012full}
B.~P. Day, A.~R. Margetts, D.~W. Bliss, and P.~Schniter, ``Full-duplex {MIMO}
  relaying: Achievable rates under limited dynamic range,'' \emph{{IEEE} J.
  Sel. Areas Commun.}, vol.~30, no.~8, pp. 1541--1553, Sep. 2012.

\bibitem{biglieri2007mimo}
E.~Biglieri, R.~Calderbank, A.~Constantinides, A.~Goldsmith, A.~Paulraj, and
  H.~V. Poor, \emph{MIMO wireless communications}.\hskip 1em plus 0.5em minus
  0.4em\relax Cambridge University Press, 2007.

\end{thebibliography}

\vfill

\end{document}